\documentclass[letterpaper, 10 pt, conference]{ieeeconf}  

\IEEEoverridecommandlockouts                              

\usepackage{graphicx} 
\usepackage{epsfig} 
\usepackage{mathptmx} 
\usepackage{times} 
\usepackage{amsmath} 
\usepackage{amssymb}  
\usepackage{arydshln} 
\usepackage[backend=bibtex,style=ieee]{biblatex}

\title{\LARGE \bf
A tutorial on the synthesis and validation of a closed-loop wind farm controller using a steady-state surrogate model
}

\author{Bart M Doekemeijer$^{\dagger}$, Paul A Fleming$^{\ddagger}$ and Jan-Willem van Wingerden$^{\dagger}$
\thanks{$^{\dagger}$Bart Doekemeijer and Jan-Willem van Wingerden are with the Delft Center for Systems and Control (DCSC), Faculty of Materials, Mechanical, and Maritime Engineering (3mE), Delft University of Technology, The Netherlands
        {\tt\small b.m.doekemeijer@tudelft.nl}, {\tt\small j.w.vanwingerden@tudelft.nl}}%
\thanks{$^{\ddagger}$Paul Fleming is with the National Renewable Energy Laboratory (NREL), Golden, Colorado, United States of America
        {\tt\small paul.fleming@nrel.gov}}%
}

\begin{document}

\maketitle
\thispagestyle{empty}
\pagestyle{empty}

\begin{abstract}
In wind farms, wake interaction leads to losses in power capture and accelerated structural degradation when compared to freestanding turbines. One method to reduce wake losses is by misaligning the rotor with the incoming flow using its yaw actuator, thereby laterally deflecting the wake away from downstream turbines. However, this demands an accurate and computationally tractable model of the wind farm dynamics. This problem calls for a closed-loop solution. This tutorial paper fills the scientific gap by demonstrating the full closed-loop controller synthesis cycle using a steady-state surrogate model. Furthermore, a novel, computationally efficient and modular communication interface is presented that enables researchers to straight-forwardly test their control algorithms in large-eddy simulations. High-fidelity simulations of a 9-turbine farm show a power production increase of up to $11\%$ using the proposed closed-loop controller compared to traditional, greedy wind farm operation.\\ 

\end{abstract}

\section{INTRODUCTION}
As wind turbines extract energy from the air stream, a slower, more turbulent flow trials behind their rotors, called the ``wake''. In wind farms, wake interaction leads to losses in power capture and accelerated structural degradation when compared to freestanding turbines (e.g., \cite{Kanev2018}). For example, for the Lillgrund offshore wind farm, wake losses have been estimated at $23\%$ in the literature \cite{Barthelmie2010}. The area of wind farm control aims to minimize these wake losses by intelligently operating the turbines in the farm. A popular method to reduce wake losses in the literature is by misaligning the rotor planes with the incoming flow using their yaw actuators, thereby laterally deflecting the wake away from downstream turbines \cite{Fleming2013}. This methodology is called ``wake redirection control'' or ``yaw control''. However, an accurate model of the wind farm is a prerequisite to accurately determine the optimal misalignment angles of the turbines \cite{Boersma2017b}.\footnote{There is research towards model-free methods for wind farm optimization (e.g., \cite{Marden2013,Rotea2017,Rotea2017b}), but the time delays involved in wake propagation pose a real challenge to such methods. This is not further explored here, and the interested reader is referred to Boersma et al. \cite{Boersma2017b}.}

\begin{figure*}
	\centering
	\includegraphics[trim=23mm 36mm 9mm 21mm,clip,width=.82\linewidth]{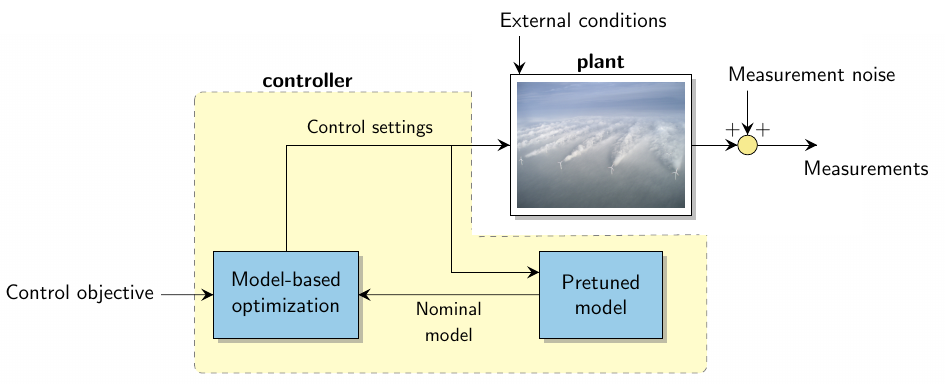}
	\caption{The open-loop control framework, in which a pretuned surrogate model is used to determine an optimal control policy, according to the assigned control objective (e.g., power maximization, power reference signal tracking).}
	\label{fig:OL_framework}
\end{figure*}

The concept of wake redirection control has been demonstrated successfully in a number of situations in the literature. Among others, \cite{Gebraad2016,Munters2017} demonstrated the concept in high-fidelity simulation. Furthermore, \cite{Campagnolo2016,Bastankhah2016} demonstrated the concept of wake redirection control in wind tunnel experiments, and \cite{Fleming2017,Fleming2017b} even tested the concept situationally in full-scale field experiments. However, all these experiments followed an open-loop approach, in which the information flows as demonstrated in Fig.~\ref{fig:OL_framework}.

\begin{figure*}
	\centering
	\includegraphics[trim=23mm 36mm 9mm 21mm,clip,width=.82\linewidth]{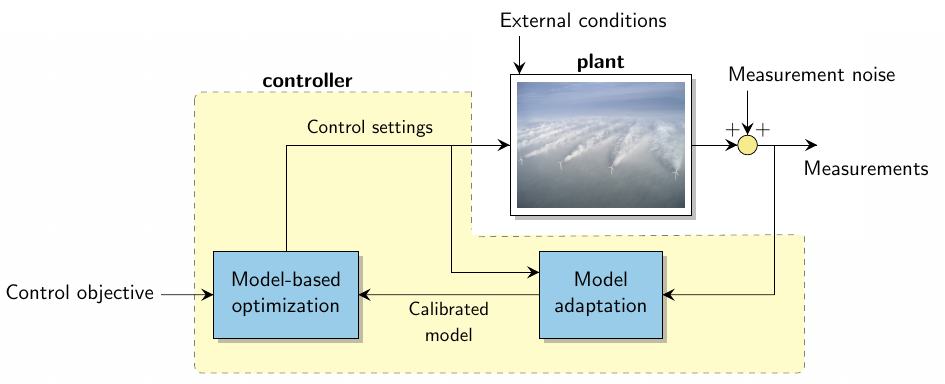}
	\caption{The closed-loop control framework, in which measurements are used in real-time to calibrate a surrogate model. This surrogate model is then used to determine an optimal control policy, according to the assigned control objective (e.g., power maximization, power reference signal tracking).}
	\label{fig:CL_framework}
\end{figure*}

However, due to the lack of information and the complicated dynamics at a range of spatial and temporal scales inside the wind farm, accurate control cannot be achieved without feedback \cite{Boersma2017b}. More precisely, the surrogate models used in the framework of Fig.~\ref{fig:OL_framework} are only accurate in particular situations, and do not suffice for all the conditions relevant throughout the annual operation of the wind farm. Hence, a closed-loop framework is preferred, in which information flows as demonstrated in Fig.~\ref{fig:CL_framework}. In this closed-loop control setting, measurements are used in a real-time optimization framework to determine the next control policy.
Here, a simplified surrogate model of the wind farm is calibrated in real-time using noisy measurements from wind farm. These measurements may originate from, e.g., sensors inside the wind turbine, measurement towers, or lidar systems \cite{Scholbrock2016}. After calibration, the surrogate model should more accurately capture the current dynamics inside the wind farm. Then, a model-based optimization algorithm employs this surrogate model to find the optimal control settings for each turbine, where optimality is defined according to the respective control objective. The frequency at which the closed-loop controller operates depends on the surrogate model, the frequency at which measurements are available, the computational hardware, and the algorithms internal to the wind farm controller (the adaptation and optimization algorithm, respectively).

\begin{figure*}[ht]
	\centering
	\includegraphics[trim=15mm 170mm 75mm 25mm,clip,width=\linewidth]{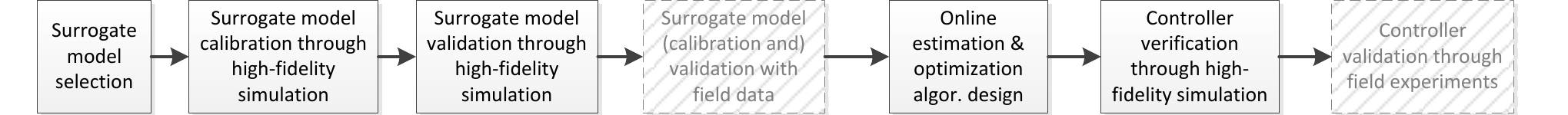}
	\caption{Flowchart for closed-loop controller synthesis}
	\label{fig:controllersynthesis}
\end{figure*}

To synthesize a closed-loop controller, a number of key steps are taken, as displayed in Fig.~\ref{fig:controllersynthesis}.
These steps are, in logical order:\\
\begin{enumerate}
	\item Surrogate model selection: the closed-loop framework of Fig.~\ref{fig:CL_framework} requires an accurate yet computationally tractable mathematical model of the dynamics inside the wind farm that are relevant for control. This can be either a steady-state model of the wind farm which predicts the time-averaged effects of a control policy on the power output of the wind farm (e.g., \cite{Jensen1983,Ainslie1988,Gebraad2016,Bastankhah2016}), but can also be a dynamic model which predicts the second-to-second flow and wind turbine dynamics (e.g., \cite{Soleimanzadeh2014,Gebraad2015,Annoni2016b,Boersma2017,Munters2017}). The surrogate model for the closed-loop controller in this work is described in Section~\ref{sec:model}.\\
	
	\item Surrogate model calibration through high-fidelity simulation: typically, surrogate wind farm models contain a number of tuning parameters which are dependent on the wind turbines and wind farm topology modeled. To push the accuracy of the surrogate model, the tuning parameters are optimized through high-fidelity simulation prior to controller algorithm design. Typically, quantities of interest to fit the surrogate model for are the turbine power capture, as this often has an important role in the optimization objective of the wind farm controller, and possibly flow dynamics at particular locations, as these may have an important contribution for the real-time calibration algorithm in the closed-loop controller (e.g., as in \cite{Bottasso2018}). A priori (offline) calibration with high-fidelity data is advantageous compared to calibration with experimental data, in the sense that measurement errors are not an issue. Furthermore, the full three-dimensional flow field is available at any point in time. In this work, the high-fidelity simulation model will be described in Section~\ref{sec:SOWFA}, after which offline model calibration will be discussed in Section~\ref{sec:calibration}.\\
	
	\item Surrogate model validation through high-fidelity simulation: Once the model has been calibrated, it's accuracy should be validated to ensure the model parameters have not been over-fit for the calibration dataset. If successful, the next step is model validation with experimental data. Model validation through simulation will be the topic of Section~\ref{sec:calibration}.\\
	
	\item Surrogate model (calibration and) validation with experimental data: after offline model calibration, the surrogate model should be validated, ideally with field data of the wind farm for which the closed-loop controller is synthesized. For example, Schreiber and Bottasso \cite{Schreiber2017} and Annoni et al. \cite{Annoni2018} have demonstrated this validation procedure for simplified steady-state surrogate models. This step is considered to be out of the scope of this paper for the presented case study.\\
	
	\item Online estimation \& optimization algorithm design:
	Once the surrogate model is validated, the controller is to be synthesized. As shown in Fig.~\ref{fig:CL_framework}, the closed-loop controller consists of two components: an estimation algorithm which adapts the surrogate model to the current conditions inside the farm in real time, and an optimization algorithm that determines the optimal control policy of the wind turbines for the conditions at hand.
	
	Literature on online estimation for wind farm surrogate models is scarce. Fortunately, additional sensing equipment in the wind farm such as lidar systems are becoming increasingly popular in the literature (e.g., \cite{Rettenmeier2014,Scholbrock2016}). This additional information may be used on the turbine level for load reduction, but can additionally be used on the wind farm level for real-time surrogate model calibration. However, currently, the step of estimation has conveniently been ignored in most of the literature on wind farm control (e.g., \cite{Munters2017,Meyers2015,Vali2017}), yielding an open-loop control solution. In some cases, a simplified state estimation algorithm has been applied for dynamic surrogate models, such as a linear Kalman filter (e.g., \cite{Iungo2015,Gebraad2015}). More recently, there have been positive developments in the field of real-time model adaptation, using more sophisticated estimation algorithms that attempt to balance accuracy with computational efficiency (e.g., \cite{Shapiro2017,Doekemeijer2018}).
	
	In terms of optimization, for steady-state surrogate models, a gradient-based or nonlinear optimization algorithm is typically employed to determine the optimal steady-state control settings for the wind farm (e.g., \cite{Gebraad2016,Thomas2017,Rott2018}). For dynamic surrogate models, typically predictive control methods are followed to yield an optimal control policy, which typically is a time-varying solution (e.g., \cite{Soleimanzadeh2012,Meyers2015,Siniscalchi2018}).
	
	Model-based estimation and optimization will be the topic of Section~\ref{sec:synthesis}.\\
	
	\item Controller verification through high-fidelity simulation: before deploying the controller in the field, it should be tested in high-fidelity simulation to ensure robustness and to provide an insight of the potential gains. Another important factor to investigate is the change in loads on the turbine structure due to the new control policy. In simulation, identical inflows can be simulated, allowing one-to-one comparisons of the controller with the baseline situation. While many controllers in the literature have been tested in simulation, they were typically assessed in idealistic conditions, using simplified models \cite{Boersma2017b}. There are only a handful of closed-loop control algorithms that were tested in a high-fidelity wind farm simulation (e.g., \cite{Gebraad2015b,Munters2017,Wingerden2017,Boersma2018}). An important contribution of this work is the facilitation of a communication infrastructure that enables researchers to test their control algorithms more easily in a high-fidelity environment. Controller verification through simulation is the topic of Section~\ref{sec:results}.\\
	
	\item Controller validation through field experiments: finally, the wind farm controller should be deployed in the field. The literature on this topic has been very limited (e.g., \cite{Fleming2017,Fleming2017b}). Generally, it may be difficult to reliably measure the gains of the closed-loop control algorithm compared to greedy control, as the ambient conditions vary continuously. Due to measurement uncertainty, the need for additional sensors and processing equipment, and the changing atmospheric conditions, controller validation through field experiments is significantly more complicated than in simulation, yet very necessary. Controller validation through field experiments is out of the scope of this work.\\
\end{enumerate}

Even though the flowchart is drawn linearly in Fig.~\ref{fig:controllersynthesis}, one has to note that it is often necessary to go through multiple iterations of simulation, algorithm development, and experimental validation, before satisfactory wind farm control performance has been realized.

Most of the literature on wind farm control has focused on only one component of the closed-loop controller synthesis for wind farms. Typically, this is either the surrogate model or the optimization algorithm. Furthermore, these solutions are typically only tested in a simplified simulation environment, and therefore the usefulness of these control solutions remains uncertain. To address this scientific gap in the literature, the main contributions of this paper are:

\begin{itemize}
	\item the explanation and demonstration of the full closed-loop controller synthesis cycle for wind farms using a steady-state surrogate model of the dominant wind farm dynamics,
	\item the development of an open-source, open-access communication interface that enables researchers to straight-forwardly test their control algorithms (developed in Python, MATLAB, or a similar language) with the high-fidelity large-eddy simulator SOWFA \cite{Churchfield2012},
	\item providing a benchmark/example simulation case in which a closed-loop wind farm control algorithm relying on a simplified surrogate model is tested in high-fidelity simulation \cite{acc2019case}.
\end{itemize}

The structure of the paper is as follows. First, a steady-state surrogate model of the wind farm is outlined in Section~\ref{sec:model}. A high-fidelity simulation model used for model calibration, validation, and controller verification is discussed in Section~\ref{sec:SOWFA}. Then, the surrogate model is calibrated and validated through high-fidelity simulation in Section~\ref{sec:calibration}. Further, a closed-loop controller is synthesized using this surrogate model in Section~\ref{sec:synthesis}. Finally, controller verification through high-fidelity simulation is the topic of Section~\ref{sec:results}. The paper is concluded in Section~\ref{sec:conclusions}.\\

\section{SURROGATE MODEL}
\label{sec:model}
The surrogate model used in this work combines a single wake model for wake redirection and turbine derating based on \cite{Bastankhah2016} with a wake deficit summation model \cite{Katic1986}, a turbine-induced turbulence model \cite{Crespo1996}, and a turbulence summation model \cite{Niayifar2016} from the literature. The focus in this section is on the single wake model, as it is the most insightful for the remainder of this paper. This surrogate model is selected for its strong theoretical origin, its performance when compared to experimental data from wind tunnel testing \cite{Bastankhah2016} and experimental data from the field \cite{Annoni2018}, and for the fact that it has fewer tuning parameters than some comparable models (e.g., \cite{Gebraad2016}). Note that this surrogate model is interchangeably called the ``FLO Redirection and Induction in Steady-state'' (FLORIS) model in this paper, and has also been published in the public domain under the same name \cite{FLORIS2018,Annoni2018}.

\begin{figure}[ht]
	\centering
	\includegraphics[width=\linewidth]{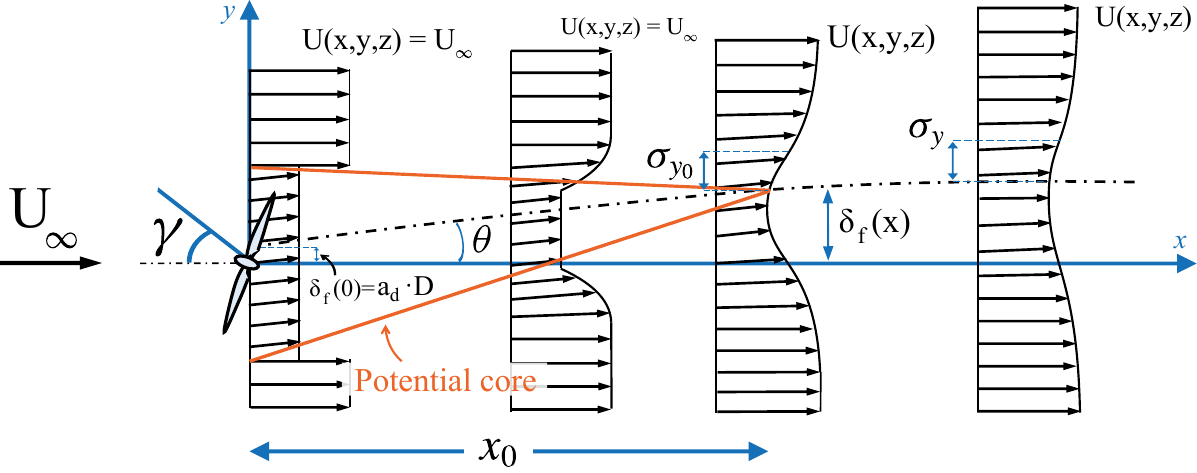}
	\caption{A schematic drawing of the single wake model, taken and modified from \cite{Bastankhah2016}.}
	\label{fig:wakemodel}
\end{figure}

In short, the near-wake zone is modeled as a linearly converging cone with its base at the turbine rotor, and its tip located at distance $x_0$ downstream. Here, $x_0$ is calculated by
\begin{equation}
\frac{x_0}{D} = \frac{\cos \left( \gamma \right) \cdot \left[ 1+ \sqrt{1-C_T} \right] }{\sqrt{2} \cdot \left(\alpha \cdot I_{\text{rotor}} + \beta \cdot \left[ 1 - \sqrt{1-C_T} \right] \right)},
\end{equation}
with $D$ the rotor diameter, $\gamma$ the yaw misalignment angle of the rotor with the incoming flow, $C_T$ the non-dimensional thrust coefficient, $I_{\text{rotor}}$ the turbulence intensity at the rotor of the turbine, and $\alpha$ and $\beta$ tuning parameters. A schematic overview of the wake model is given in Fig.~\ref{fig:wakemodel}.

At the onset of the near wake and in the far wake region, the wake deficit follows the shape of a two-dimensional Gaussian distribution, according to
\begin{align}
\begin{split}
\frac{U(x,y,z)}{U_\infty} = 1 - &\left(1-\sqrt{1- \frac{\sigma_{y0} \sigma_{z0}}{\sigma_y \sigma_z} C_T }\right) \times \\
&\text{exp} \left( \frac{(y-\delta_f)^2}{2 \sigma_y^2} + \frac{z^2}{2 \sigma_z^2} \right),
\end{split}
\end{align}
where $U_\infty$ is the wind speed far upstream of the turbine, $\sigma$ is the standard deviation in the specified direction, and $(x,y,z)$ is the Eucledian space with its origin at the turbine hub, $x$ aligned with the wind direction and $z$ being positive upwards. The Gaussian-shaped wake is centralized around the centerline. The centerline is displaced in $y$-direction from the $x$-axis by distance $\delta_f$ due to a yaw misalignment and the rotor rotation, calculated as
\begin{align}
\begin{split}
\delta_f =& \delta_r(x) + \tan \left( \theta \right) x_0 + \frac{\theta}{5.2} \cdot \left(C_0^2 - 3e^{1/12}C_0 + 3 e^{1/3} \right) \times \\ & \sqrt{\frac{\sigma_{y0} \sigma_{z0}}{ k_y \cdot k_z \cdot C_T }} \cdot \ln \left[ \frac{\left(1.6+\sqrt{C_T}\right) \left( 1.6 S_{\sigma} - \sqrt{C_T} \right) }{ \left(1.6-\sqrt{C_T} \right) \left( 1.6 S_{\sigma} + \sqrt{C_T} \right)} \right].
\end{split}
\end{align}
Here, $\theta$ is the initial deflection angle, calculated as
\begin{equation}
\theta \approx \frac{0.3 \gamma }{ \cos \gamma } \left(1- \sqrt{1-C_T \cos \gamma} \right).
\end{equation}
Furthermore, $C_0 = 1-\sqrt{1-C_T}$, $k_y$ and $k_z$ are linear wake expansion coefficients similar to that in Jensen \cite{Jensen1983}, and $S_\sigma$ is defined as $
S_\sigma = \sqrt{(\sigma_y \sigma_z)/(\sigma_{y0} \sigma_{z0})}$, with $\sigma_y$ and $\sigma_z$ the standard deviations of the Gaussian in the y- and z-direction, respectively. These are calculated as
\begin{align}
\sigma_y &= \sigma_{y0} + (x-x_0) k_y, ~~ \text{ with } ~~ \sigma_{y0} = \frac{D}{2\sqrt{2}} \cos \gamma,\\
\sigma_z &= \sigma_{z0} + (x-x_0) k_z, ~~ \text{ with } ~~ \sigma_{z0} = \frac{D}{2\sqrt{2}}.
\end{align}
The wake expansion coefficients are a function of $I_{\text{rotor}}$, as
\begin{equation}
k_y = k_z = k_a \cdot I_{\text{rotor}} + k_b,
\end{equation}
with $k_a$ and $k_b$ tuning parameters. Further, $\delta_r$ is the wake deflection induced by the rotation of the blades, approximated using a linear function following the idea of Gebraad et al. \cite{Gebraad2016}, by $\delta_r = a_d \cdot D + b_d \cdot x$, with $a_d$ and $b_d$ tuning parameters.

Finally, the time-averaged power capture of a turbine 
is calculated by combining the effects of all wakes impinging this turbine's rotor following Katic \cite{Katic1986} and actuator disk theory. The interested reader is referred to related literature \cite{Bastankhah2016,Annoni2018} for more information. \\ 

\section{HIGH-FIDELITY MODEL}
\label{sec:SOWFA}
In this work, the Simulator fOr Wind Farm Applications (SOWFA), a high-fidelity simulation model from the U.S. National Renewable Energy Laboratory (NREL), is used for model calibration, model validation and controller verification \cite{Churchfield2012,sowfa}.\\

\subsection{SOWFA}
SOWFA is a large-eddy simulation model that incorporates a rotating actuator disk implementation of wind turbines, and solves the three-dimensional, filtered, unsteady Navier-Stokes equations over a finite temporal and spatial mesh, accounting for Coriolis and geostrophic forcing terms. Large-eddy simulation models such as SOWFA resolve larger scale flow dynamics directly, and employ a subgrid-scale model for smaller eddy dynamics. SOWFA has been used on multiple occasions for surrogate model calibration (e.g., \cite{Gebraad2016,Annoni2016a,Boersma2017}), model validation, and wind farm controller verification (e.g., \cite{Doekemeijer2018,Gebraad2015b,Annoni2016a,Wingerden2017}).\\

\subsection{Wind farm controller interface}
In order to test controllers in a closed-loop setting, measurement data and control settings need to be passed between SOWFA and an external wind farm controller periodically throughout the simulation. As most wind farm control algorithms from the literature are implemented in Python or MATLAB, and SOWFA operates in C, coupling these pieces of software is not straight-forward. For this reason, an important contribution of this work is the development of an interface that allows researchers to test their control algorithms with SOWFA without making significant modifications to their code.

The open-source software zeroMQ \cite{zeromq} was implemented in SOWFA as a message passing interface to an external wind farm controller and published in the public domain \cite{sowfa}. Using this interface, one can straight-forwardly expand their wind farm controller implemented in a programming language of choice (supporting zeroMQ) to receive measurement data from SOWFA, and return control settings. Note that SOWFA and the wind farm controller are run in parallel, and can even operate on different computers, platforms, and networks -- as long as a (network-based) connection can be made. Currently, the TCP protocol is used for communication.

\begin{figure}[ht]
	\centering
	\includegraphics[trim=25mm 20mm 185mm 40mm,clip,width=.98\linewidth]{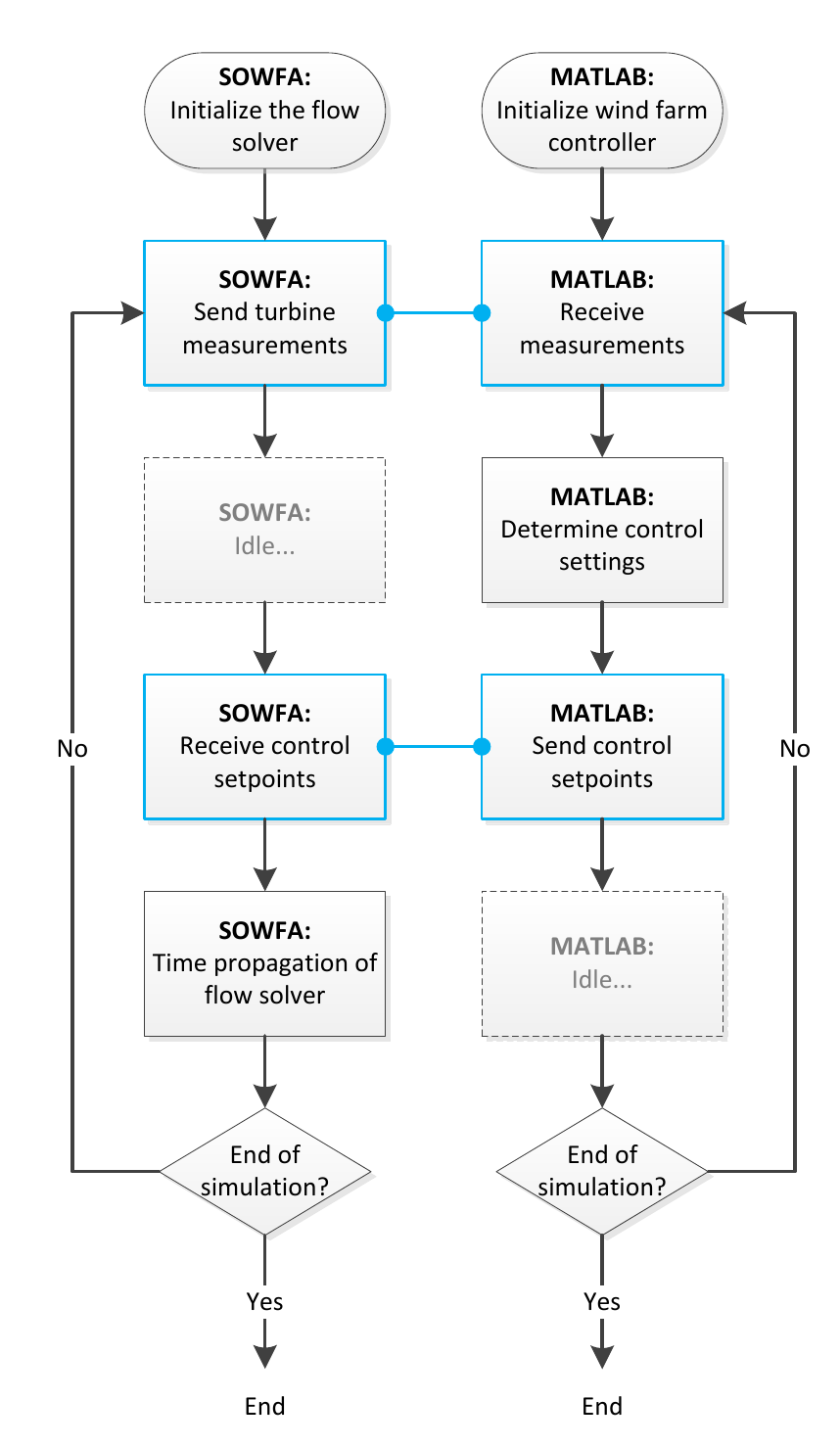}
	\caption{This figure shows a flowchart of the order of operations in a SOWFA simulation which is coupled to a closed-loop wind farm controller (in this case: implemented in MATLAB).}
	\label{fig:flowchart_zmq}
\end{figure}

The order of operations in a typical wind farm simulation using the ZeroMQ interface is shown in Fig.~\ref{fig:flowchart_zmq}.
In this case, MATLAB is used as an example in which the wind farm control algorithm is implemented. Note that SOWFA and MATLAB are run in parallel, rather than in serial. After initialization, each waits for the other to perform its computations, and thus only one of the two is really performing computations at any point in time. Hence, the idea is to have SOWFA and MATLAB share the same computational cores to minimize the time that cores spend idling. Communication through ZeroMQ happens twice each discrete timestep of the simulation -- once to transmit measurements to MATLAB, and once to receive control settings from MATLAB.\\

\section{MODEL CALIBRATION \& VALIDATION THROUGH HIGH-FIDELITY SIMULATION}
\label{sec:calibration}
The model presented in Section~\ref{sec:model} has a number of tuning parameters that may vary with, e.g., the wind farm topology and wind turbine types. Specifically, some of the literature of Section~\ref{sec:model} is based on wind tunnel experiments, in which flow behavior is known to deviate from the actual large-scale wind farms. For this reason, the parameters $\alpha$, $\beta$, $k_a$, $k_b$, $a_b$ and $b_d$ are tuned to high-fidelity, true-scale wind farm simulation data in this section, prior to control algorithm design.

We perform a set of single-turbine simulations in SOWFA to calibrate the surrogate model with. This set contains:
\begin{itemize}
	\item Two types of inflow: one set with uniform inflow and one set with turbulent inflow,
	\begin{itemize}
		\item each consisting of a set of simulations with yaw angles ranging from $-30^{\circ}$ to $30^{\circ}$ with a collective blade pitch angle of $0^{\circ}$, 
		\item and another set of simulations with a turbine yaw angle of $0^{\circ}$ and the collective blade pitch angle varying from $1^{\circ}$ to $4^{\circ}$.
	\end{itemize}
\end{itemize}  This covers both wake deflection and turbine derating for typical turbine operation. The NREL 5MW turbine is used \cite{Jonkman2005}. Using this data, the model is now calibrated offline as follows.\\
\subsection{Calibration methodology}
\begin{enumerate}
	\item A spatially and temporally averaged vertical inflow profile is extracted from the high-fidelity dataset. The same inflow profile is used in the surrogate model through a linear spline interpolation. An example is given in Fig.~\ref{fig:comparison_inflow}.
	\begin{figure}[ht]
	\centering
	\includegraphics[width=6.3cm]{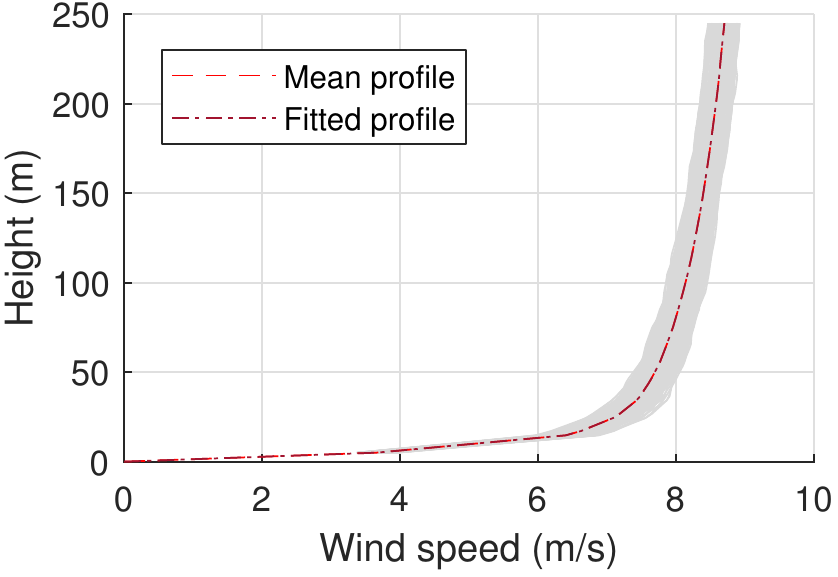}
	\caption{Inflow comparison. In gray are all vertical profiles along the spatial domain, upon which a single mean profile is fit using spline interpolation.}
	\label{fig:comparison_inflow}
	\end{figure}

	\item The flow field from the high-fidelity simulation is time-averaged over a 10-minute window to average local fluctuations. This fits the scope of what the surrogate model intends to reproduce.
	\item This time-averaged flow field is sliced at $3D$, $5D$, $7D$ and $10D$ downstream, and measurements are sampled over a rectangular grid at each downstream location. An example is shown in Fig.~\ref{fig:comparison_5D} for one of the simulations with uniform inflow.
	
	\begin{figure}[ht]
		\centering
		\includegraphics[width=.7\linewidth]{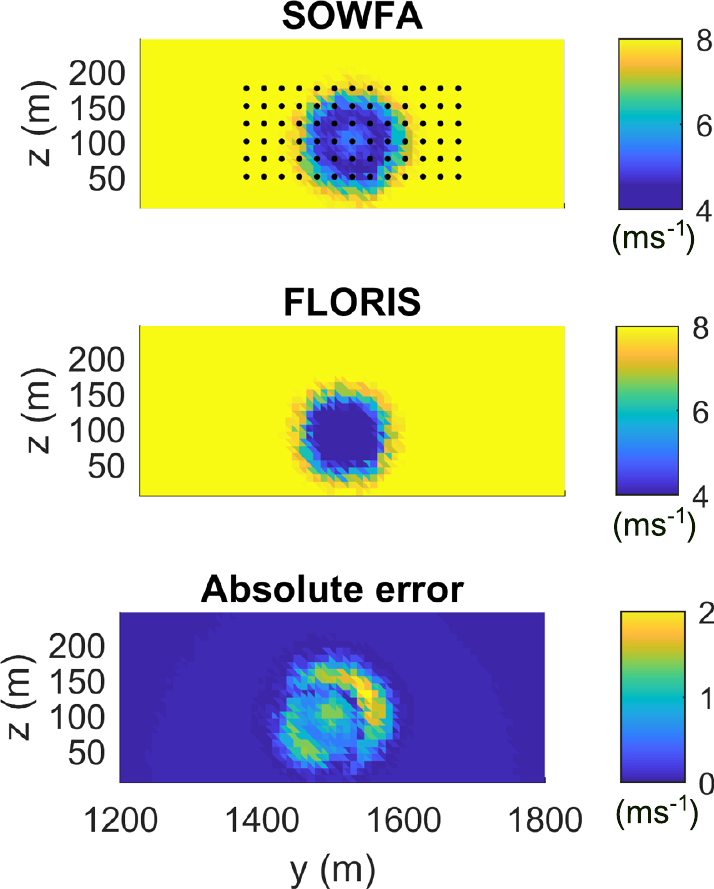}
		\caption{Wake comparison at 5D downstream. The black dots in the top subplot show the locations of the measurements that will be used to calibrate the surrogate model with SOWFA. 
		}
		\label{fig:comparison_5D}
	\end{figure}

	\item A cost function is set-up, where the root-mean-squared error between the flow measurements from SOWFA and that predicted by the surrogate model is minimized for arguments $\Psi = \begin{bmatrix} \alpha & \beta & k_a & k_b & a_d & b_d 	\end{bmatrix}$, as:
	\begin{equation}
	\Psi_{\text{opt}} = \arg \min_{\Psi} \sum_{i} \left( U_{\text{SOWFA}}^i - U_{\text{FLORIS}}(\Psi) \right),
	\label{eq:optimization}
	\end{equation}
	where $i$ covers the full set of single-turbine simulations. The control settings and ambient conditions varying with $i$ are neglected in notation here.
	
\end{enumerate}

An important remark is that, in a more elaborate study, one would have to include the combined effect of turbine derating and wake redirection. Furthermore, a wider range of turbulent inflows should be considered, at various turbulence intensities and various mean wind speeds. Also, it is important to consider the interaction for multiple turbine wakes. However, this is outside of the scope of this work.\\

\subsection{Calibration results}
The model described in Section~\ref{sec:model} has been implemented in MATLAB, and made available to the public \cite{FLORIS2018}. A constrained genetic algorithm optimization approach is used to solve the problem of Eq.~\ref{eq:optimization} in an efficient, parallelized manner, taking approximately $20$~CPU-hours. The optimized parameters $\Psi_{\text{opt}}$ are displayed in Table~\ref{tab:model-parameters}. The lower and upper bounds on the parameter optimization space are chosen as to stay within the same order of magnitude as the nominal values presented in the literature \cite{Bastankhah2016}, in order to limit overfitting and parameter divergence.

\begin{table}[ht]
	\centering
	\caption{Optimal parameters $\Psi_{\text{opt}}$ for the surrogate model after calibration using high-fidelity simulation data}
	\label{tab:model-parameters}
	\begin{tabular}{l l l l}
		Variable & Lower bound & Upper bound & Optimal value \\ \hline
		$\alpha$&	$~~5.80 \cdot 10^{-1}$ & $~~9.28$ & $~~3.16$ \\
		$\beta$ &	$~~3.85 \cdot 10^{-2}$ & $~~6.16 \cdot 10^{-1}$ & $~~3.28 \cdot 10^{-1}$ \\
		$k_a$ & 	$~~9.59 \cdot 10^{-2}$ & $~~1.53 \cdot 10^{0}$ & $~~1.74 \cdot 10^{-1}$ \\
		$k_b$ & 	$~~9.25 \cdot 10^{-4}$ & $~~1.48 \cdot 10^{-2}$ & $~~9.69 \cdot 10^{-4}$ \\
		$a_d$ & 	$-1.00  $ & $ ~~1.00 $ & $-1.34 \cdot 10^{-3}$ \\
		$b_d$ & $-4.00 \cdot 10^{-2}$ & $-2.50 \cdot 10^{-3}$ & $-2.68 \cdot 10^{-3}$
	\end{tabular}
\end{table}

Inspecting Table~\ref{tab:model-parameters}, it is seen that most of the optimized values lay between their lower and upper bound. This is a good sign, as the opposite situation may indicate overfitting and parameter divergence.\\

\subsection{Validation results}
\begin{figure}[ht]
	\centering
	\includegraphics[width=.675\linewidth]{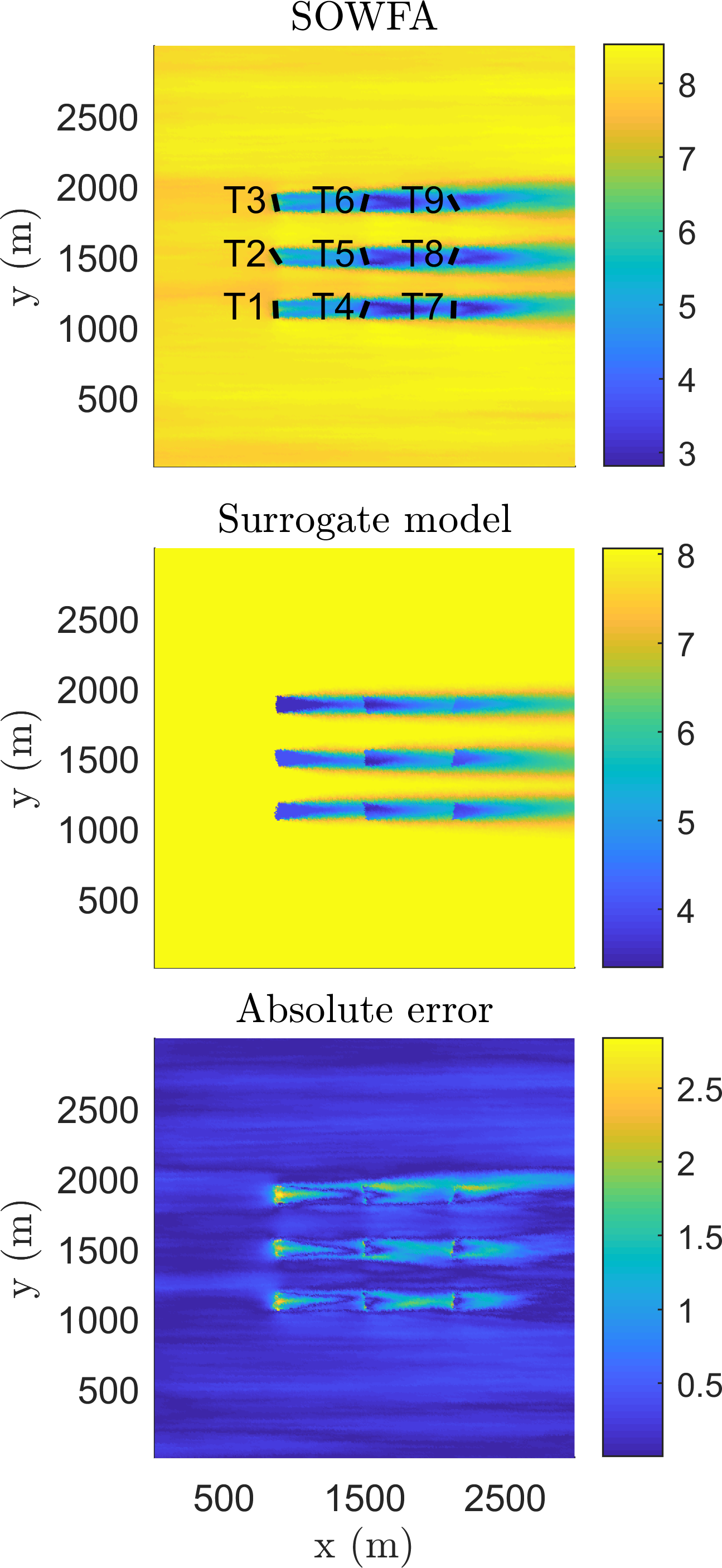}
	\caption{Validation of the surrogate model with SOWFA: time-averaged flow field at turbine hub height. Units are ms$^{-1}$.}
	\label{fig:validation_flow}
\end{figure}

\begin{figure}[ht]
	\centering
	\includegraphics[width=\linewidth]{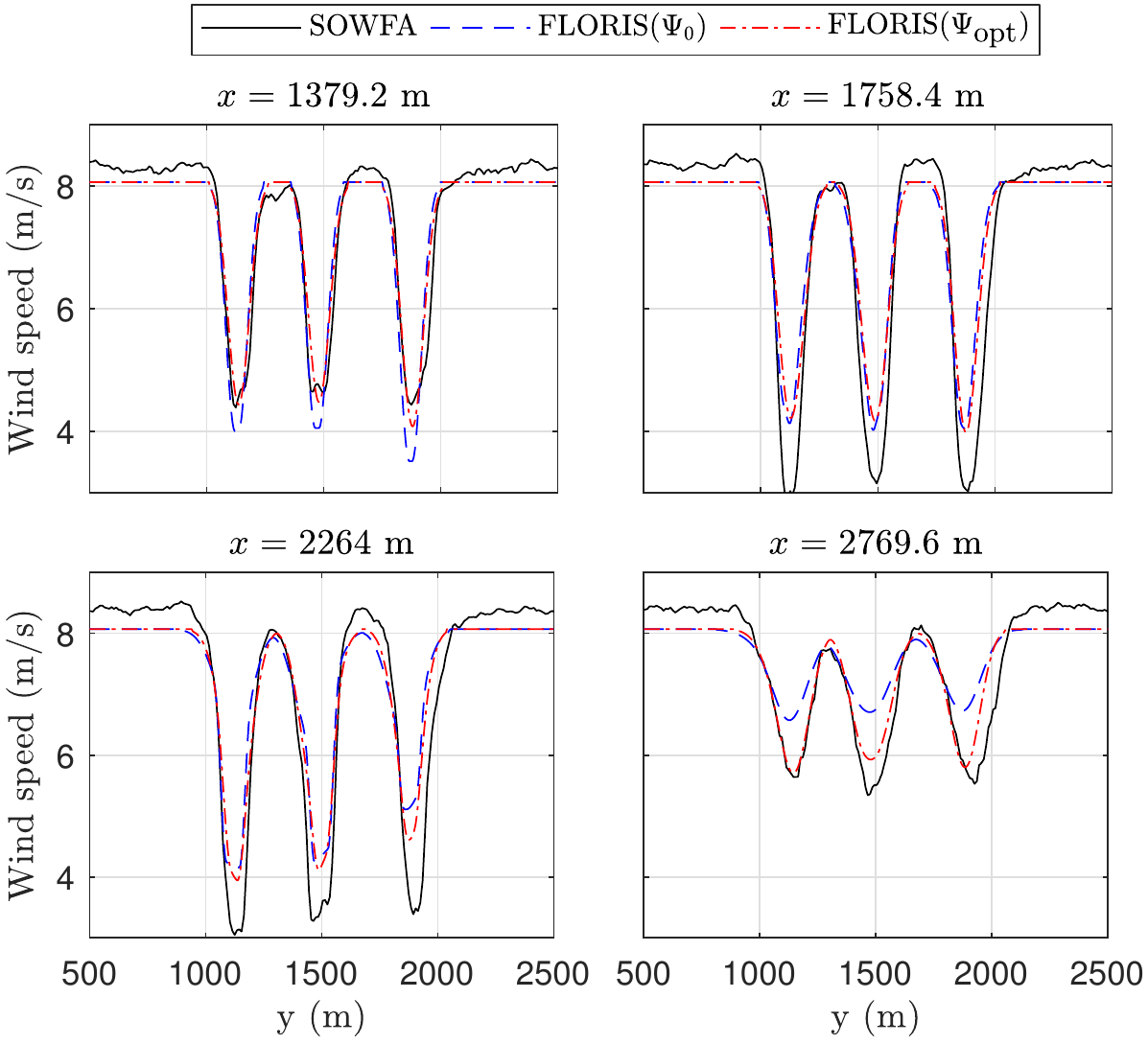}
	\caption{Validation of the surrogate model with SOWFA: wind speed at hub height at different distances downstream}
	\label{fig:validation_wakedeficit}
\end{figure}

To ensure that the model calibration procedure was successful, the calibrated model is compared with a high-fidelity simulation dataset of a 9-turbine wind farm in which arbitrary yaw angles are applied to the turbines. The model has not been fit for wake interaction, and hence this is an interesting case to inspect. The yaw angles are derived from a Gaussian distribution, yielding 
\begin{align*}
\vec{\gamma} = [&2.9^{\circ},~ 32.1^{\circ},~   12.6^{\circ},~ -20.3^{\circ},~   16.1^{\circ},~  \\
 &-14.4^{\circ},~   -1.9^{\circ},~  -21.6^{\circ},~   29.4^{\circ}].
\end{align*}
Note that the pitch angles are kept constant at $0^{\circ}$ in this validation case, since it is in unlikely that they will be exploited for wind-farm-wide power maximization in the to-be-synthesized optimization algorithm \cite{Annoni2016a}. The time-averaged horizontal plane is shown in Fig.~\ref{fig:validation_flow}. Furthermore, the time-averaged wake deficits at different distances downstream are displayed in Fig.~\ref{fig:validation_wakedeficit}.
From these figures, a good fit can be seen in the far-wake regions and in the single-turbine wakes. As more wakes interact, the fit gets worse, as the model has not been calibrated for this situation. Furthermore, the calibrated model parameters $\Psi_{\text{opt}}$ have improved the model compared to the nominal model parameters from the literature $\Psi_0$, especially for the near-wake and far-downstream region.

\begin{figure}[ht]
	\centering
	\includegraphics[width=.9\linewidth]{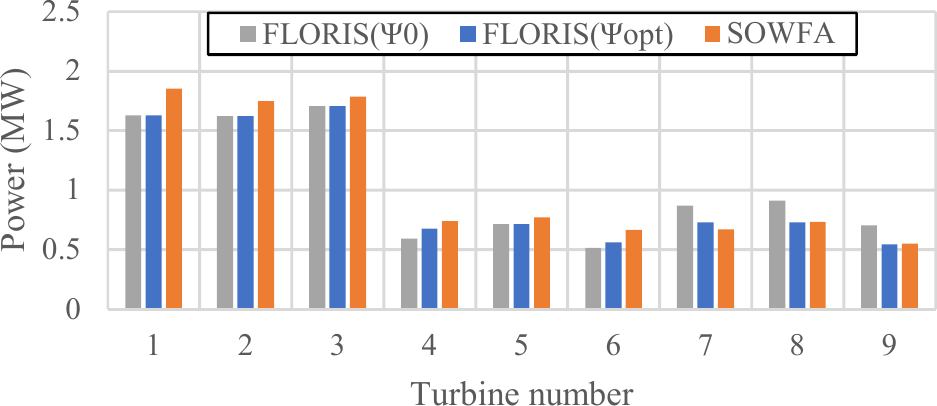}
	\caption{Validation of the surrogate model with SOWFA: time-averaged power capture per turbine}
	\label{fig:powervalidation}
\end{figure}

The power predicted from the surrogate model is compared to the power from SOWFA in Fig.~\ref{fig:powervalidation}.
One can see that the trends are adequately captured in the surrogate model. Though, it slightly underestimates the power capture in most situations. Furthermore, the calibrated parameters $\Psi_{\text{opt}}$ show improved performance compared to $\Psi_0$.

In conclusion, the surrogate model can accurately capture the wake and power of this 9-turbine wind farm. However, it is still to be seen whether the surrogate model can capture more difficult situations such as deep-wake effects and partial overlap situations such as described in \cite{Martinez2018}. This should be addressed in future work.\\

\section{CLOSED-LOOP CONTROLLER SYNTHESIS}
\label{sec:synthesis}
The turbine control settings inside the wind farm are optimized using the surrogate model from Section~\ref{sec:model} in a closed-loop setting. The model was calibrated on $10$-minute average data in Section~\ref{sec:calibration}. In the proposed closed-loop controller, the control settings of the turbines are optimized every $10$ minutes. The controller consists of two components: an estimator and an optimizer, each described next. \\

\subsection{Estimation}
In this work, online estimation is limited to the ambient conditions: the freestream wind direction, turbulence intensity and mean wind speed. A single-shot estimation of all three ambient parameters using only turbine power measurements would most likely result in parameter divergence and overfitting. For example, in a two-turbine case, one can almost always bring the cost function to zero by choosing a certain (wrong) wind direction and wind speed.

To avoid overfitting, firstly the wind direction is assumed to be estimated for each turbine individually following the approach of \cite{Bertele2017}. Secondly, the wind speed and turbulence intensity are collectively estimated on a farm-wide level by minimizing a weighted root-mean-squared error of the measured and predicted turbine power signals of each turbine, putting a higher weight on the upstream turbines, as
\begin{equation}
\Xi_{\text{opt}}  = \arg \min_\Xi \sum_{i}^{N_t} w_i \left( P^i_{\text{SOWFA}} - P^i_{\text{FLORIS}}(\Xi) \right).
\label{eq:estimation}
\end{equation}
Here, $N_t$ indicates the number of turbines, $w_i$ is a weighing term, and $\Xi = [I_\infty, U_{\infty}]$ is to be estimated.\\


\subsection{Optimization}
\begin{figure}[ht]
	\centering
	\includegraphics[width=.95\linewidth]{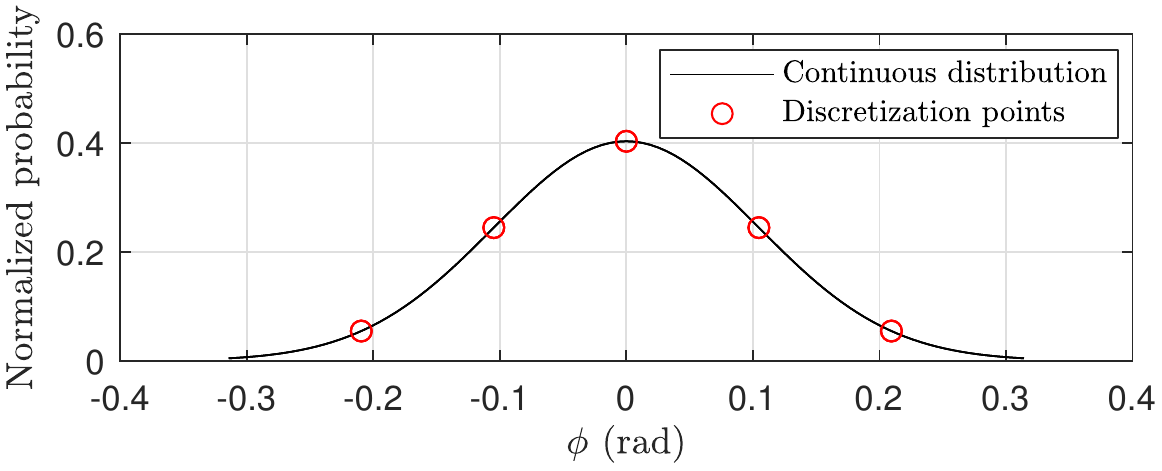}
	\caption{Robust optimization}
	\label{fig:robustoptim}
\end{figure}

The wind direction is known to have a significant impact on the optimal yaw angles inside the wind farm \cite{Rott2018}. The approach used to estimate the wind direction was derived from \cite{Bertele2017}. In the corresponding paper, a standard deviation in wind direction estimation of $6^{\circ}$ was given for $U_\infty = 8$~m/s. Hence, a robust optimization approach is followed. In this case, we use the approach from Rott et al. \cite{Rott2018}, in which the yaw angles are optimized for a probability distribution of wind directions, rather than one deterministic wind direction. The optimization is formulated as follows,
\begin{equation}
\vec{\gamma}_{\text{opt}} = \arg \max_{\vec{\gamma}} \int_{-\pi}^{\pi} \rho(\phi) \sum_{i}^{N_t} P^i_{\text{FLORIS}}(\gamma_i),
\label{eq:robustoptimization}
\end{equation}
with $\vec{\gamma} = \begin{bmatrix} \gamma_1 & \gamma_2 & \cdots & \gamma_{N_t} \end{bmatrix}$, $N_t$ the number of turbines, and $\rho$ a probability distribution of variable $\phi$, the wind direction. Basically, the yaw angles are now optimal if they provide consistent performance for a range of wind directions in proximity of the mean wind direction. For a solution to exist within a reasonable computational time, the probability distribution is discretized at 5 points, as demonstrated in Fig.~\ref{fig:robustoptim}.
Thus, for one function evaluation, the surrogate model is simulated five times, each with a different wind direction. In the example case of Fig.~\ref{fig:robustoptim}, this would be $\phi=-0.21$~rad, $\phi=-0.10$~rad, $\phi=0$~rad, $\phi=0.10$~rad, and $\phi=0.21$~rad. The resulting farm-wide power capture for each of these five simulations are summed, weighted according to their respective probability. The objective is to maximize this weighted sum. Note that in terms of measurement uncertainty, the standard deviation of $\rho$ goes down with the square root of the number of individual sensors. In this case, the number of individual sensors is equal to $N_t$, as each turbine is assumed to provide a unique measurement of (what is assumed to be) the same quantity.\\

\section{SIMULATION RESULTS}
\label{sec:results}
The 9-turbine wind farm from Fig.~\ref{fig:validation_flow} is used to test the closed-loop controller described in Section~\ref{sec:synthesis}. The turbines are initialized at a greedy control setting, where $\gamma_i = 0^{\circ} ~\forall~ i$. Then every $600$~s, the controller determines the optimal yaw angles, constrained with $-25^{\circ} \le \gamma_i \le 25^{\circ} ~ \forall ~ i$ to threshold the increase in structural loads due to a yaw misalignment.\footnote{Note that pitch angles were also optimized by the controller, but were found to be zero in all cases.}\\ 

\subsection{Open-loop controller}
In the first case, an open-loop (OL) controller is synthesized where a single set of time-invariant ambient conditions $\Xi$ is assumed. Specifically, $\Xi$ contains the freestream wind direction $\phi$, the turbulence intensity $I_{\infty}$ and the mean ambient wind speed $U_\infty$, respectively. The true ambient conditions are $\Xi_{\text{true}} = \begin{bmatrix}
0.0^{\circ} & 6.0\% & 8.0~\text{ms}^{-1}
\end{bmatrix}$. In the open-loop case, we simulate the situation of a model mismatch by assuming $\Xi = \begin{bmatrix} 10.0^{\circ} & 1.0\% & 6.5~\text{ms}^{-1} \end{bmatrix}$. Additionally, in the OL controller, the probability distribution $\rho$ in Eq.~\ref{eq:robustoptimization} is assumed to have a zero standard deviation. 
The optimal control settings for the OL controller are displayed in Table~\ref{tab:OL-results}.

\begin{table}[ht]
	\centering
	\caption{Optimal control policy using open-loop control}
	\label{tab:OL-results}
	\begin{tabular}{l r r r r r}
		Time [s] & $\vec{\gamma}_{\text{opt},1:2}$[$^{\circ}$] & $\vec{\gamma}_{\text{opt},3}$[$^{\circ}$] & $\vec{\gamma}_{\text{opt},4}$[$^{\circ}$] & $\vec{\gamma}_{\text{opt},5:6}$[$^{\circ}$] & $\vec{\gamma}_{\text{opt},7:9}$[$^{\circ}$] \\\hline
		$0$ & $0.0$ & $0.0$ & $0.0$ & $0.0$ & $0.0$ \\
		$600$ & $-9.6$ & $-10.9$ & $-12.0$ & $-11.0$ & $0.0$ \\
		$1200$ & $-9.6$ & $-10.9$ & $-12.0$ & $-11.0$ & $0.0$ \\
		$1800$ & $-9.6$ & $-10.9$ & $-12.0$ & $-11.0$ & $0.0$ \\
	\end{tabular}
\end{table}
As this is a steady-state, open-loop controller, the optimal yaw angles are time-invariant throughout the simulation.\\

\subsection{Closed-loop controller}
In the second case, two closed-loop (CL) controllers are synthesized, in which the past $300$~s of measurements are time-averaged and used to estimate the ambient conditions $\Xi$. The weights $w_i$ in Eq.~\ref{eq:estimation} are set to be $3$ for upstream turbines $w_{1,2,3}=3$, the weights are $w_{4,5,6}=2$ for the first row of downstream turbines, and $w_{7,8,9}=1$ for the most downstream turbines. Basically, it is assumed that the confidence is highest with the unwaked turbines, and the model fit is progressively worse with more wake interactions.

The two closed-loop controllers differ in their optimization approach. The standard deviation of $\rho$ in Eq.~\ref{eq:robustoptimization} is assumed to be $0^{\circ}$ for the deterministic CL controller, and $\frac{6}{\sqrt{9}} = 2^{\circ}$ for the robust CL controller, as there are nine turbines providing a unique measurement of (what is assumed to be) the same quantity. The resulting control policy for the robust CL controller is shown in Table~\ref{tab:CL-results}.

\begin{figure}[ht]
	\centering
	\includegraphics[width=\linewidth]{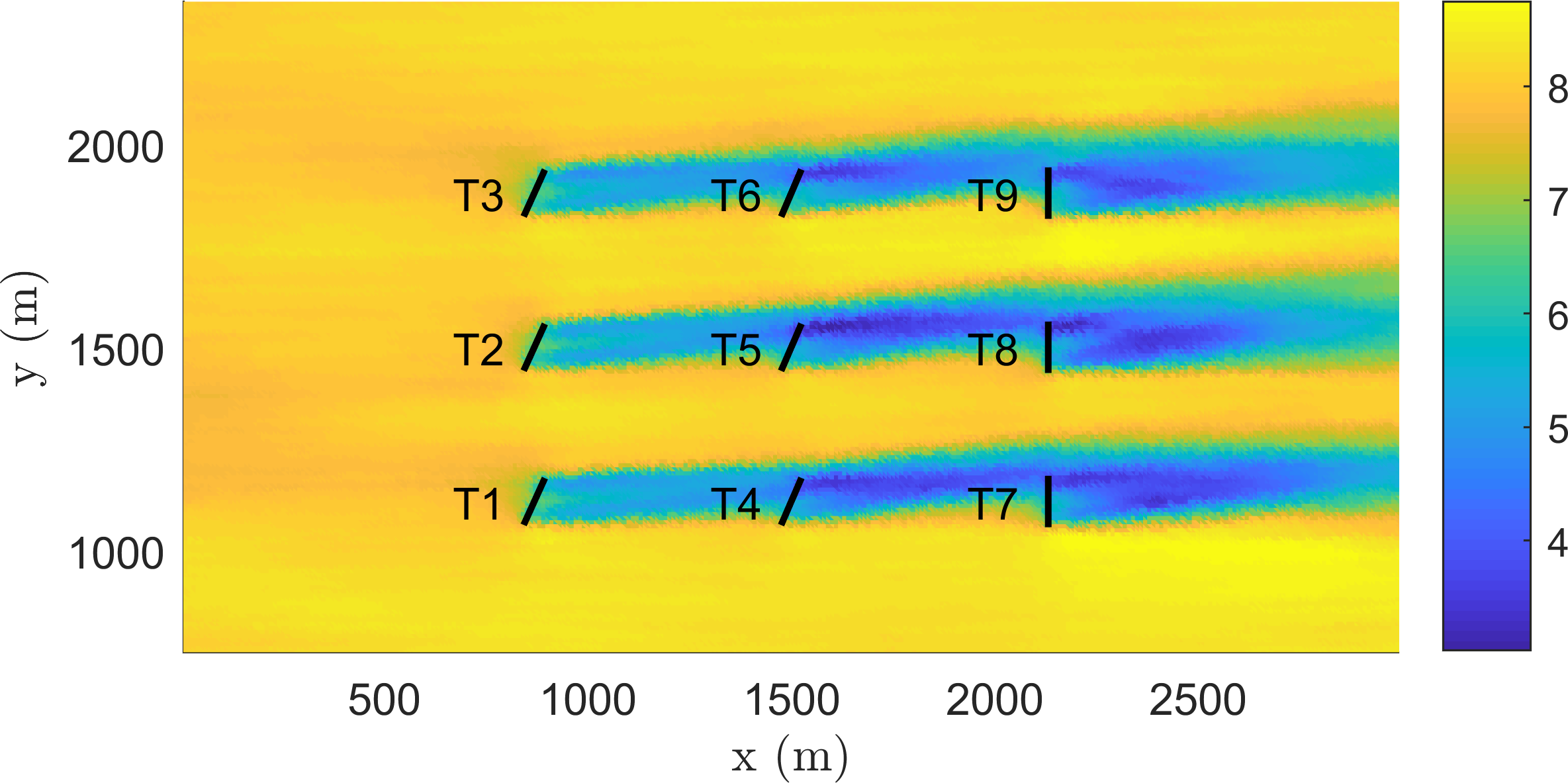}
	\caption{Time-averaged wind speed in ms$^{-1}$ at the horizontal flow slice at $z=90$~m (hub-height) for $t=900$ to $t=1200$~s.}
	\label{fig:SSC_900-1200s}
\end{figure}

\begin{table}[ht]
	\centering
	\caption{Optimal control policy using closed-loop control and a robust optimization methodology}
	\label{tab:CL-results}
	\begin{tabular}{l l r r r}
		Time [s] & $\Xi_{\text{opt}}$ [$^{\circ}$, $\%$, ms$^{-1}$] & $\vec{\gamma}_{\text{opt},1:3}$[$^{\circ}$] & $\vec{\gamma}_{\text{opt},4:6}$[$^{\circ}$] & $\vec{\gamma}_{\text{opt},7:9}$[$^{\circ}$] \\\hline
		$0$ & N/A & $0.0$ & $0.0$ & $0.0$ \\
		$600$ & $\begin{bmatrix} \hspace{2.1mm} 2.3 & 6.0 & 8.16 \end{bmatrix}$ & $-24.4$ & $-23.3$ & $0.0$ \\
		$1200$ & $\begin{bmatrix} -3.7 & 7.6 & 8.15 \end{bmatrix}$ & $19.6$ & $18.3$ & $0.0$ \\
		$1800$ & $\begin{bmatrix} \hspace{2.1mm} 1.1 & 6.9  & 8.15 \end{bmatrix}$ & $ 24.5$ & $18.7$ & $0.0$
	\end{tabular}
\end{table}
The optimal yaw angles provided by the robust CL controller vary with time, due to the changing atmospheric conditions $\Xi$. Note that there is some switching behavior at $t=1200$~s in the optimal yaw angles due to the change in sign of the estimated wind direction. While the robust optimization approach should account for this \cite{Rott2018}, it is expected that the standard deviation for $\rho$ in the optimization was too low. This should be investigated in future work. The time-averaged flow field in SOWFA under the closed-loop control policy for $t=900$~s to $t=1200$~s is shown in Fig.~\ref{fig:SSC_900-1200s}.\\

\subsection{Comparison}
The performance of the open-loop and closed-loop controllers is shown in Table~\ref{tab:OL_CL_comparison}.
\begin{table}[ht]
	\centering
	\caption{Comparison of greedy case, open-loop controller case, and the closed-loop controllers cases}
	\label{tab:OL_CL_comparison}
	\begin{tabular}{l r r r r r}
		Time window & $\gamma = \vec{\gamma}_{\text{greedy}}$ & $\gamma = \vec{\gamma}^{\text{~det.}}_{\text{opt,~OL}}$ & $\gamma = \vec{\gamma}^{\text{~det.}}_{\text{opt,~CL}}$ & $\gamma = \vec{\gamma}^{\text{~rob.}}_{\text{opt,~CL}}$ \\\hline
		0-600 s & 10.71 MW & 10.71 MW & 10.71 MW & 10.71 MW \\ 
		600-900 s & 9.78 MW & 9.71 MW & 9.29 MW & 9.40 MW \\
		900-1200 s & 9.49 MW & 9.80 MW & 10.44 MW & 10.49 MW \\
		1200-1500 s & 9.54 MW & 9.78 MW & 10.50 MW & 10.45 MW \\
		1500-1800 s & 9.64 MW & 9.92 MW & 10.32 MW & 10.34 MW \\ \hdashline 
		0-2000 s & 10.07 MW & 10.21 MW & 10.40 MW &10.45 MW 
	\end{tabular}
\end{table}
In this table, $\vec{\gamma}^{\text{~det.}}_{\text{opt,~CL}}$ and $\vec{\gamma}^{\text{~rob.}}_{\text{opt,~CL}}$ are the sets of optimal yaw angles obtained in closed-loop by optimizing with a standard deviation for the ambient wind direction $\rho$ in Eq.~\ref{eq:robustoptimization} of $0^{\circ}$ and $2^{\circ}$, respectively. One can see that both the OL and the CL controllers improve the farm-wide power capture compared to a greedy control approach. However, the CL controllers consistently outperform the OL controller, with a situational wind-farm-wide power increase of approximately $3\%$ for the OL controller compared to greedy wind farm operation, and between $7\%$ and $11\%$ for the CL controllers. This is due to the fact that the surrogate model more accurately captures the current conditions inside the farm for the CL controllers. The only loss compared to greedy control is for $t=600-900$~s, in which the effect of yawing the upstream turbines has not yet resulted in a weaker wake on the downstream rotors. Furthermore, the robust optimization approach leads to slightly better performance when compared to an approach in which the wind direction is assumed to be deterministic.

\begin{figure}[ht]
	\centering
	\includegraphics[width=\linewidth]{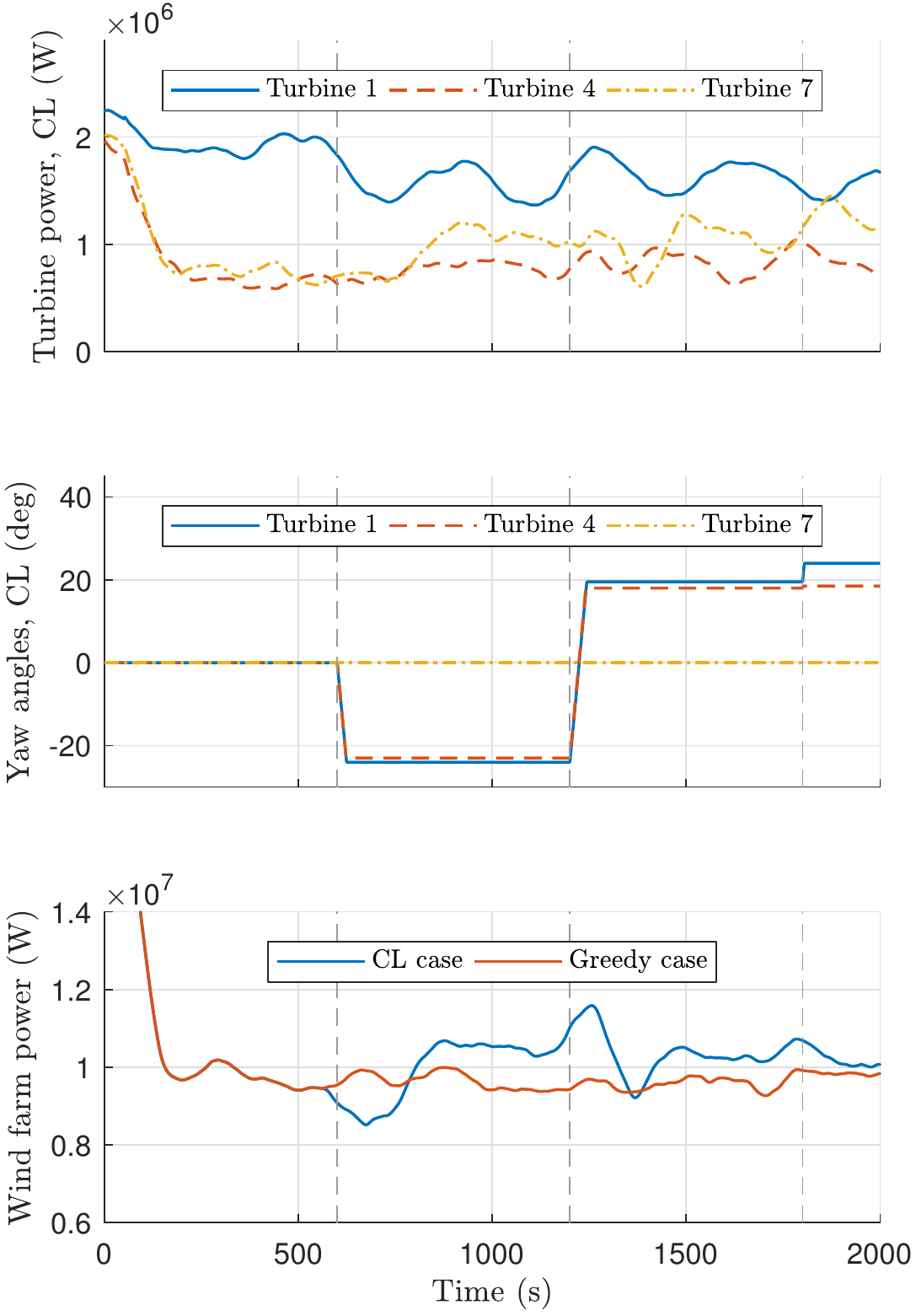}
	\caption{Timeseries of the power capture and turbine yaw angles for the robust closed-loop wind farm simulation. The first and third subplot show moving averages of the turbine resp. wind farm power capture, averaged by a time window of $50$~s in both the past and future data. The yaw angles are updated according to the controller at a rate of $600$~s, starting from the traditional greedy control strategy at $0$~s.}
	\label{fig:powerTimeseries}
\end{figure}

The loss in power capture at $t=600-900$~s can be more explained using Fig.~\ref{fig:powerTimeseries}. 
In this figure, the power signals are time-averaged with a moving average filter (non-causal low-pass filter) with a time window of $50$~s for both past and future data, to provide more insight. As turbines 1 and 4 are purposely misaligned with the incoming flow at $600$~s, they see a loss in power capture shortly after $600$~s. While this generates weaker wakes behind turbines 1 and 4, it takes some time for these weaker wakes to propagate to turbines 4 and 7 downstream. Once that happens, a significant gain can be seen (especially for turbine 7, around $800$~s). Furthermore, a similar effect can be seen at $1200$~s, since wind turbines 1 and 4 are now yawed towards the opposite direction. This leads to a temporary increase in power capture, as the turbines pass $\gamma=0^{\circ}$, but eventually also to a decrease due to the stronger wakes generated downstream. After the flow settles, there is again a constant gain in overall power capture compared to the greedy control case.\\

\subsection{Discussion}
While the closed-loop controllers presented in this section yield a significant increase in wind-farm-wide power production compared to traditional, greedy operation, an important remark should be made. Namely, the controllers have been simulated under idealistic ambient conditions. While the large-eddy SOWFA simulation is of significantly higher fidelity than the surrogate model, the inflow in SOWFA still only has one mean wind direction, wind speed, and turbulence intensity. As the control settings are optimized at a frequency of every $10$ minutes, this may or may not be fast enough in a more realistic setting where ambient conditions slowly vary with time. The actuation frequency necessary for wind farm control remains an open question in the research \cite{Boersma2017b}. Because of this, the fidelity necessary for surrogate models also remains an open question, and a wide range of steady-state and dynamical models is investigated in the literature. The study of SCADA data, experimental testing and high-fidelity, large-eddy simulations coupled with mesoscale models should further provide guidance in answering these questions. This is out of the scope of this work.\\ 

\subsection{Wind farm controller interface}
An important contribution of this work is the open-source communication interface developed for the verification of wind farm control algorithms in a high-fidelity environment. For the simulations presented in this section, the computational time required by the ZeroMQ communication was found to be on the order of $1 \cdot 10^{-3}$~s for sending a single message (set of measurements or control settings) in either direction. This is negligible compared to the computational cost of SOWFA, which is in the order of $10^1$~s per timestep for this parallelized 80-core case.\\

\section{CONCLUSIONS}
\label{sec:conclusions}
This paper demonstrated the synthesis cycle of a closed-loop wind farm controller using a steady-state surrogate model. The surrogate model was first calibrated and validated using high-fidelity simulations, after which the controller was tested in a high-fidelity 9-turbine wind farm simulation. To facilitate the testing of wind farm controllers written in MATLAB or Python in a high-fidelity environment, a communication interface was developed for the high-fidelity simulator SOWFA. SOWFA simulations with closed-loop wind farm control showed an increase in wind farm power capture of $7$\% to $11$\% through yaw control. Furthermore, the proposed communication architecture has a negligible computational cost. While positive results were shown in the simulations presented in this work, the ambient conditions vary slowly in real wind farms. In theory, this closed-loop framework should be able to deal with such changes. This should be explored further in future work.\\


\section*{ACKNOWLEDGMENT}
The authors would like to thank Manon Kok for the insightful discussions on real-time model calibration of the surrogate model. 
This project has received funding from the European Union's Horizon 2020 research and innovation programme under grant agreement No 727477.\\

\section*{SOFTWARE AVAILABILITY}
All software presented in this work is open-source. The high-fidelity simulation software SOWFA is developed by NREL \cite{sowfa}. This software repository also includes the communication interface that exchanges information through ZeroMQ. The surrogate model presented in this work is actively being developed by the Delft University of Technology \cite{FLORIS2018}. The community-driven ZeroMQ library is also available in the public domain \cite{zeromq}. The wind farm controller relying on the surrogate model, exchanging information through the ZeroMQ interface with SOWFA, has also been made available \cite{acc2019case} as an example and a benchmark case.\\

\printbibliography

\end{document}